\documentclass{PoS}

\usepackage{graphicx}
\usepackage[square,sort&compress]{natbib}
\usepackage{overpic}
\usepackage{amsfonts, amsmath, amsthm, amssymb} 
\usepackage{wasysym}
\usepackage[amssymb,textstyle]{SIunits}
\usepackage{booktabs}
\usepackage{afterpage}

\addunit{\GeV}{\giga\electronvolt}
\addunit{\TeV}{\tera\electronvolt}
\addunit{\eV}{\electronvolt}
\addunit{\smm}{\square\metre\usk\second}
\addunit{\smmr}{\rpsquare\metre\usk\reciprocal\second}

\newcommand{\e}[1]{\cdot\power{10}{#1}}

\usepackage{wrapfig} 

\abstract{
The cores of ultra-luminous infrared galaxies (ULIRGs) are very dense environments, with a high rate of star formation and supernova explosions. They are thought to be sites of cosmic-ray acceleration, and are predicted to emit $\gamma$-rays in the \GeV{} to \TeV{} range. So far, no ULIRG has been detected in $\gamma$-rays. Arp 220, the closest ULIRG to Earth, has been well studied, and detailed models of $\gamma$-ray production in this galaxy are available. They predict a rather hard $\gamma$-ray spectrum up to several \TeV{}. Due to its large rate of star formation, high gas density, and its close proximity to Earth, Arp 220 is thought to be a very good candidate for observations in very-high-energy (VHE; \unit{100}{\GeV} -- \unit{100}{\TeV}) $\gamma$-rays. Arp 220 was observed by the VERITAS telescopes for more than 30 hours with no significant excess over the cosmic-ray background. The upper limits on the VHE $\gamma$-ray flux of Arp 220 derived from these observations are the most sensitive limits presented so far and are starting to constrain theoretical models. We also present upper limits for the VHE flux from the ULIRG IRAS 17204-0014, the starburst galaxy IC 342, and the active galaxy 3C321.
}

\FullConference{The 34th International Cosmic Ray Conference,\\
		30 July -- 6 August, 2015\\
		The Hague, The Netherlands}

\ShortTitle{Upper limits on the VHE $\gamma$-ray flux from the ULIRG Arp 220 and other galaxies}

\author{\speaker{Henrike Fleischhack} ~for the VERITAS collaboration\thanks{\texttt{http://veritas.sao.arizona.edu}} \\
        DESY , Platanenallee 6, 15738 Zeuthen, Germany\\
        E-mail: \email{henrike.fleischhack@desy.de}}

\title{Upper limits on the VHE $\gamma$-ray flux from the ULIRG Arp 220 and other galaxies with VERITAS }

\begin{document}

\section{Starburst Galaxies}
Star formation rates vary from one galaxy to another, as well as for the same galaxy over time. Galaxies with a high rate of star formation (compared to other galaxies, or to the same galaxy at other times) are called \textit{starburst galaxies} (SBGs). This period of high star formation generally only lasts for a small fraction of the lifetime of the galaxy. Starbursts may be triggered by interactions with other galaxies, e.g., mergers. Studying these types of galaxies can provide insight into the evolution of galaxies in general as well as our own Milky Way.

SBGs show strong emission over many wavelengths. For example, there is a well-studied correlation between the emission in radio and far infrared (FIR) that is seen in samples of star forming galaxies \citet{LTQ}, thought to be driven by the star forming rate. A high star formation rate means lots of young, massive, stars, which radiate mostly in the UV and have short lifespans on the order of tens of millions of years. The UV emission is absorbed by the interstellar dust and re-emitted in the FIR. At the end of their lives, many massive stars explode as supernovae. Supernova remnants (SNRs) can accelerate electrons and positrons to high speeds, causing them to emit synchrotron radiation in radio wavelengths. 
Thus, galaxies with a high star formation rate have high luminosities both in radio and FIR. Galaxies with a FIR luminosity $L^\mathrm{FIR}$ of more than $10^{12}\cdot L_{\astrosun}$ are also known as ultra-luminous infrared galaxies.

Similar to the radio-FIR correlation, the existence of a correlation between HE (high energy, \unit{100}{\mega\electronvolt{}} -- \unit{100}{\GeV{}}) and VHE $\gamma$-ray emission and FIR emission has been suggested as well \citet{LTQ}. Again, this correlation is thought to be driven by the star formation rate. Young SNRs can accelerate cosmic rays, which produce $\gamma$-rays by interaction with the ISM. $\gamma$-rays are either produced by the decay of neutral pions, or by cosmic ray electrons (including secondary electrons from pion decays) that upscatter IR or radio photons to higher energies. Note that this correlation only strictly holds in the \textit{calorimetric limit}, when the galaxy is dense enough that most cosmic rays interact, rather than leave the galaxy \citet{fermiarpUL,m82ngc253}.

Only eight star forming galaxies, four of which are part of the local group, have been detected by the \textit{Fermi}-LAT \citet{fermiarpUL}, at energies of \unit{0.1}{\GeV} to \unit{100}{\GeV}. The authors of \citet{fermiarpUL} show that for their sample, the HE luminosities were well correlated with the FIR luminosities. None of the non-detections of other star forming galaxies seem to contradict this correlation.

Only two star forming galaxies have been detected in VHE $\gamma$ rays: M82 \citet{m82nature} and NGC 253 \citet{ngc253hess}, which is too few to study correlations. However, assuming the VHE-FIR correlation holds, we can use the VERITAS measurement of the $\gamma$-ray flux from M82 to estimate the fluxes from other star forming galaxies. Given the luminosity distance $D_\mathrm{L}$, the \TeV{} flux $F^\mathrm{TeV}$ of a star forming galaxy can be estimated as

\begin{equation*}
F^\mathrm{TeV} = \frac{L^\mathrm{FIR}}{D_L^2} \cdot \frac{D_{L,\mathrm{M82}}^2 \cdot F^\mathrm{TeV}_\mathrm{M82}}{L_\mathrm{M82}^\mathrm{FIR}} .
\end{equation*}

This has been calculated for the three SBGs under study here, see Table \ref{tblPred}. Note that these estimates are very rough. There have been several efforts to model the cosmic ray acceleration, interaction of cosmic rays with the ISM, and production of VHE $\gamma$-rays in either single galaxies \citet{somecomments} or a larger sample of star forming galaxies \citet{LTQ}. For example, Table \ref{tblArp220} shows the predictions for the VHE $\gamma$-ray flux from Arp 220 by some of those more detailed models. The flux predictions range from one order of magnitude below to one order of magnitude above the naive prediction. EBL absorption as well as absorption inside the source galaxy are neglected here as they are not important for photons with energies below a few \TeV{} for the galaxies studied here.
 
\begin{table}[tb]
\centering
\begin{tabular*}{0.95\textwidth}{@{\extracolsep{\fill} } lrrrr}
\toprule\toprule
\textbf{Source} & $\mathbf{D_L}$ & $\mathbf{L_\mathrm{8-1000\mu m}}$ & $\mathbf{E_{\mathrm{min}}}$  & \textbf{Flux} $\mathbf{(E>E_\mathrm{min})}$ \\ 
 & [kpc] & [$\power{10}{10}\cdot L_{\astrosun}$] & [\GeV] & [\smmr] \\ 
\midrule
M82 \citet{m82nature}	& 3.4	& 4.6	&	700	&	$3.7\e{-9}$	 \\ 
\midrule
IC342				& 3.7	& 1.4	&	500	&	$\sim 1.6\e{-9}$ \\ 
Arp 220				& 74.7	& 140	&	500	&	$\sim 0.4\e{-9}$ \\ 
IRAS 17208-0014		& 173	& 230	&	600	&	$\sim 0.1\e{-9}$ \\ 
\bottomrule
\end{tabular*} 
\caption{Rough predictions for the VHE flux of three starburst galaxies. Luminosity distances $L_D$ and FIR luminosities $L_\mathrm{8-1000\mu m}$ are taken from \citet{fermiarpUL}. The M82 flux as measured by VERITAS \citet{m82nature} was used as the reference VHE flux. A spectral index of $-2.5$ was assumed for all sources, consistent with the M82 measurements and the expectations for other star forming galaxies.}
\label{tblPred}
\end{table}

\subsection{The ULIRG Arp 220}
Arp 220 is the closest ULIRG to Earth, at a distance of about 75\,kpc. It is the product of a merger of two galaxies and retains two dense nuclei, sites of very high star formation. Due to its high density and close proximity to Earth, it is considered a good candidate to search for HE and VHE $\gamma$ rays from its star forming regions. Several groups have created detailed models of the cosmic-ray acceleration, propagation and interaction in Arp 220 and other galaxies, and used these models to predict the VHE $\gamma$-ray emission \citet{ArpNew,arpFirst,somecomments,LTQ}. Some of these predictions are collected in Table \ref{tblArp220}. All of the models shown here were tuned to be consistent with the measured radio emission. Due to the uncertainties in the astrophysical parameters, there are degeneracies in the model parameters that cannot be resolved using this approach. Hence, there is a significant spread between the predictions of the different modeling approaches, and also between different parameter sets for each model.  

The $\gamma$-ray telescopes H.E.S.S. and MAGIC have looked for VHE $\gamma$-ray emission from Arp 220, but neither has detected a significant excess over the background expectations. H.E.S.S. has published upper limits on the integral flux (see \citet{arphess2006,arphess2008} and Table \ref{tblArp220}). MAGIC has published upper limits on the differential flux \citet{magicarpUL} which are above even the more optimistic predictions by \citet{arpFirst,somecomments}.
\begin{table}[tb]
\centering
\begin{tabular*}{0.95\textwidth}{@{\extracolsep{\fill} } lrrr}
\toprule\toprule
\textbf{Source} & $\mathbf{E_{\mathrm{min}}}$  &  $\mathbf{E_{\mathrm{max}}}$  & \textbf{Flux} $\mathbf{(E_\mathrm{min}} $ to $\mathbf{ E_\mathrm{max})}$ \\ 
 & [\GeV] & [\GeV] & [\smmr] \\ 
\midrule
Prediction, \citet{LTQ}, increased wind, B-field & 500 & --- & * 4\e{-9} \\
Prediction, \citet{LTQ}, standard parameters & 500 & --- & * 1.5\e{-9} \\ 
Prediction, \citet{somecomments}, Blattnig pp cross section & 500 & --- & 8\e{-9} \\ 
Prediction, \citet{somecomments}, Aharonian pp cross section & 500 & --- & 2\e{-9} \\ 
Prediction, \citet{ArpNew}, most optimistic & 500 & 3000 & 2\e{-10} \\ 
Prediction, \citet{ArpNew}, most pessimistic & 500 & 3000 & 5\e{-11} \\ 
\midrule
HESS flux upper limit \citet{arphess2006} & 1000 & --- & $\leq 3.3\e{-9}$ \\ 
HESS flux upper limit \citet{arphess2008} & 1430 & --- & $\leq 2.5\e{-9}$ \\ 
VERITAS flux upper limit, this work & 500 & --- & $\leq 2.2\e{-9}$ \\ 
\bottomrule
\end{tabular*} 
\caption{Predictions and upper limits on the integral flux of Arp 220 in the VHE range. The two values marked with an asterisk (*) were interpolated using the predictions of the fluxes above \unit{300}{\GeV{}} and above \unit{1}{\TeV{}}, assuming a power-law spectrum. Note that the Blattnig parametrization of the proton-proton (pp) cross section is now thought to overestimate the VHE $\gamma$-ray flux.
}
\label{tblArp220}
\end{table}

\subsection{The ULIRG IRAS 17208-0014 and the SBG IC342 }
IRAS 17208-0014 and IC342 are two star forming galaxies selected as good candidates to search for VHE $\gamma$-ray emission. IRAS 17208-0014 is very bright in FIR and not too far from Earth, about twice the distance to Arp 220. IC342 is not nearly as luminous in FIR, but it is very close to Earth and was thus also considered as a good candidate for observations.

The Whipple 10m telescope did not detect any VHE $\gamma$ ray emission from IC 342, leading to an upper limit on the integral flux above \unit{600}{\GeV} of \unit{5.5\e{-8}}{\smmr} (at $99.9\%$ confidence) \citet{whippleIC342}.

\subsection{The Active Galaxy 3C321}
3C321, the \textit{Death Star Galaxy}, is a very peculiar system of two galaxies in the process of merging. Both have active nuclei, and the larger galaxy has a jet  pointing at its companion. 3C321 has been imaged in many wavelengths from radio to X-ray \citet{3c321maps} and there is strong evidence that the jet is interacting with the companion galaxy. The observed X-ray emission can be seen as evidence of particle acceleration at the interaction sites. Depending on the nature and energy spectrum of the accelerated particles and the surrounding medium, it might be possible for HE or VHE $\gamma$-rays to be produced as well.

There have been no published studies on $\gamma$-ray emission from 3C231 in the HE or VHE range so far.

\section{VERITAS observations}
The Very Energetic Radiation Imaging Telescope Array System (VERITAS) \citet{veritas, naheeperformance} is an array of four imaging atmospheric Cherenkov telescopes located at the Fred Lawrence Whipple Observatory (FLWO) in southern Arizona (31 40N, 110 57W,  1.3km a.s.l.). Since its inauguration in 2007, it has been upgraded several times. With the current configuration, VERITAS is sensitive to air showers induced by $\gamma$-rays with energies between \unit{85}{\GeV} and $>$\unit{30}{\TeV}. Its field of view has a diameter of \unit{3.5}{\degree}, so VERITAS operates in pointing mode: Each night, a number of sources are observed for between 15 minutes and a few hours at a time. VERITAS has a single-photon angular resolution of less than \unit{0.1}{\degree} at \unit{1}{\TeV} and an energy resolution of $15-25\%$. Its peak effective area is about \power{10}{5}\metre\squared{}. Due to the large background of cosmic-ray induced showers and the generally low fluxes at very high energies, observations over many nights have to be combined to detect weak sources. With the current configuration, VERITAS can detect a point-like $\gamma$-ray source with a flux of $1\%$ of the flux of the Crab Nebula in about 25 hours.

An overview of the data used to obtain the results presented here can be found in Table \ref{tblObs}. All data were cleaned and analyzed with two independent analysis packages \citet{veritasanalysis}, using a Hillas-based analysis. The energy reconstruction was done using lookup tables. We applied $\gamma$-hadron separation cuts that were optimized for hard-spectrum sources. All observations were carried out in \textit{wobble} mode, with the telescopes pointing \unit{0.5}{\degree} away from the source, enabling the simultaneous estimation of the remaining cosmic-ray background from off-source counts.

The two objects Arp 220 and 3C321 are separated by about \unit{1}{\degree} in the sky, and thus can be observed simultaneously with VERITAS. However, the sensitivity drops off for larger offsets. The analysis of 3C321 uses both data taken pointing at 3C321 (about \unit{10}{\hour} in 2008, with VERITAS in its original configuration) and pointing at Arp 220 (about \unit{31}{\hour} in 2012, after the first upgrade of the array). The loss in sensitivity due to pointing away from the source is offset by the larger exposure, and the increased sensitivity of the array after the first upgrade. 
For the analysis of Arp 220, only data taken pointing at Arp 220 were used, since adding a small data set with poor sensitivity does not improve our results.

\begin{table}[t]
\centering
\begin{tabular*}{0.95\textwidth}{@{\extracolsep{\fill} } lrrr}
\toprule\toprule
\textbf{Source} & \textbf{Observing period} & \textbf{Livetime} & \textbf{Mean zenith angle} \\
 & [MJD] & [\hour] & [\degree] \\
\midrule
Arp 220 & $55987 - 56046$ & 31 & 16 \\ 
IRAS 17208-0014 & $55678-56008$ & 14 & 34 \\ 
IC342 & $55528-55572$ & 3 & 37 \\ 
3C321 & $54503-56046$ & 41 & 17 \\ 
\bottomrule 
\end{tabular*} 
\caption{Overview over VERITAS data analyzed for this paper.}
\label{tblObs}
\end{table}

\section{VERITAS results}

None of the four objects considered for this analysis showed a significant excess above the background expectation. Thus, upper limits on the $\gamma$-ray flux from these objects have been calculated. Flux limits and energy thresholds can be found in Table \ref{tblResults}. 

\subsection{Upper limit procedure}
We define the ON region as a circular region in the sky with a radius of \unit{0.09}{\degree}, centered on the source position. Even after $\gamma$-hadron separation cuts, some level of background contamination is present in the ON region. We can estimate that level of background by measuring the counts in a source-free OFF region with the same acceptance as the ON region. We assume that the counts in the ON and OFF regions, $N_\mathrm{ON}$ and $N_\mathrm{OFF}$, are distributed according to Poissonian distributions with expectation values $\lambda_{ON}=N_S+N_B$ and $\lambda_{OFF}=\frac{N_B}{\alpha}$. Here, $N_S$ ($N_B$) is the expected number of signal (background) counts in the ON region, and $\alpha$ is the ratio between the area of the ON and OFF regions. 

We can now estimate the number of signal events present in the ON region as $\widehat{N_S} = N_\mathrm{ON} - \alpha \cdot N_\mathrm{OFF}$. The $\gamma$-ray flux would then be estimated as $\widehat{F} = \frac{\widehat{N_S}}{A_\mathrm{eff}\cdot T}$, where $A_\mathrm{eff}$ is the spectral-weighted effective collection area of the array (determined from simulations) and $T$ is the observing time, corrected for dead time due to detector readout.

If there is no significant excess of ON counts over the background expectation, we can instead calculate an upper limit $N_S^\mathrm{UL}$ on the number of signal events and from that, an upper limit on the integral flux via $F^\mathrm{UL} = \frac{N_S^\mathrm{UL}}{A_\mathrm{eff}\cdot T}$. The upper limits shown here are calculated following Rolke et al. \citet{trolke}, method 4 (background from sidebands, known efficiency $\tau=\frac{1}{\alpha}$) and reported at $99\%$ confidence level. 
\begin{table}[tb]
\centering
\begin{tabular*}{0.95\textwidth}{@{\extracolsep{\fill} } lrrrrrr}
\toprule\toprule
\textbf{Source} & $\mathbf{E_{\mathrm{min}}}$  & $\mathbf{N_\mathrm{ON}}$ & $\mathbf{N_\mathrm{OFF}}$ & $\mathbf{\alpha}$ & $\mathbf{A_\mathrm{eff}\cdot T}$ & \textbf{Flux} $\mathbf{(E>E_\mathrm{min})}$  \\ 
 &  [\GeV] & & & & [\smm] & [\unit{\smmr}] \\
\midrule
Arp 220			& 500  & 45 & 282 & $0.17$ & 8.35\e{9} & $\leq 2.24\e{-9}$ \\  
IRAS 17208-0014	& 600 & 16 & 124 & $0.17$ & 4.76\e{9} & $\leq 1.82\e{-9}$ \\ 
IC342 			& 500 & 37 & 174 & $0.18$ & 0.94\e{9} & $\leq 27.6\e{-9}$ \\ 
3C321 			& 500 & 52 & 363 & $0.17$ & 7.23\e{9} & $\leq 1.90\e{-9}$ \\ 
\bottomrule
\end{tabular*} 
\caption{$99\%$ confidence level upper limits on the integral flux of the four galaxies studied here, measured by VERITAS. A power law spectrum is assumed for all sources, with an index of $-3.0$ for 3C321 and index $-2.5$ for the three star forming galaxies.}
\label{tblResults}
\end{table}

\subsection{Uncertainties}
There are two sources of uncertainties that have to be considered here: our ignorance of the source spectrum and the systematic uncertainty on the VERITAS energy scale/effective area.

The effective collection area of the VERITAS instrument depends on the source spectrum. For the three starburst galaxies, it is assumed that the $\gamma$-ray spectrum should follow the cosmic ray spectrum \citet{m82ngc253,somecomments,LTQ,arpFirst}. Following \citet{arpFirst}, we assume that the spectrum can be modeled as a power law with a spectral index of $-2.0$ to $-2.5$ at very high energies. Deviations from the power law spectrum are expected, e.g., due to absorption in the host galaxy, or by extra-galactic background light, but these effects can be neglected below some \unit{10}{\TeV}. We have calculated the upper flux limits for several values of the assumed spectral index; here, we quote the weakest limits which were obtained for an assumed spectral index of $-2.5$. For 3C321, there is no published model/prediction of the VHE $\gamma$-ray spectrum. We calculated upper limits assuming a power-law spectrum with an index of $-3.0$, $-3.5$, and $-4.0$. The limits reported here are for an assumed index of $-3$, which gave us the weakest limits of the three test cases.

There is a $20\%$ systematic uncertainty on the VERITAS energy scale, which translates into a $30\%$ systematic uncertainty on the integral flux for a spectral index of $-2.5$, and a $40\%$ uncertainty on the flux for a spectral index of $-3.0$.

\section{Summary and conclusions}
Star forming galaxies, especially SBGs and (U)LIRGs, are expected to emit $\gamma$-rays in the HE and VHE range. Analogous to the radio-FIR correlation, a correlation between FIR emission and (V)HE emission has been proposed. Emission in these three regimes is thought to be driven by the star formation rate. The detection of M82 and NGC 235 in VHE by VERITAS and H.E.S.S., and of eight star forming galaxies by the \textit{Fermi}-LAT, supports these expectations. However, in order to make sure that our understanding of the underlying processes is correct, it is important to investigate a range of galaxies with different properties and star formation rates.

We have searched for VHE $\gamma$-ray emission from three star forming galaxies with VERITAS: the ULIRGs Arp 220 and IRAS 17208-0014, and the SBG IC342. No signal was detected and upper limits on the VHE flux from these galaxies were obtained. These upper limits are more constraining than previously published observations by other experiments. The non-detection is not very surprising, given our sensitivity compared to the predicted fluxes.

In the case of Arp 220, the VERITAS upper limits are starting to constrain the theoretical predictions. We can clearly exclude the more `optimistic' model by \citet{LTQ} at the $99\%$ confidence level, even considering the systematic uncertainties of the measurement. However, the upper limits presented here are still consistent with the lower range of predictions from this models. There are considerable differences between the predictions by different models. For example, a very recent model \citet{ArpNew} puts the predicted emission from Arp 220 about one magnitude lower than previous models and thus far out of reach for VERITAS.
Very deep observations by current generation IACTs would be necessary to significantly improve on those flux limits or discover VHE $\gamma$-ray emission from Arp 220. Future experiments with larger effective collection area, such as the planned CTA observatory \citet{cta}, are expected to make definite improvements and either observe the VHE $\gamma$ rays, or seriously challenge our understanding of the acceleration, propagation and interaction of cosmic rays in star forming galaxies.

\acknowledgments
VERITAS is supported by grants from the U.S. Department of Energy Office of Science, the U.S. National Science Foundation and the Smithsonian Institution, and by NSERC in Canada. We acknowledge the excellent work of the technical support staff at the Fred Lawrence Whipple Observatory and at the collaborating institutions in the construction and operation of the instrument. The VERITAS Collaboration is grateful to Trevor Weekes for his seminal contributions and leadership in the field of VHE gamma-ray astrophysics, which made this study possible. 

H.F. acknowledges support through the Helmholtz Alliance for Astroparticle Physics.

\bibliography{bib}

\providecommand{\href}[2]{#2}\begingroup\raggedright\begin{thebibliography}{10}

\bibitem{LTQ}
B.~C. Lacki, T.~A. Thompson, and E.~Quataert, {\it {The Physics of the
  Far-infrared-Radio Correlation. I. Calorimetry, Conspiracy, and
  Implications}},  {\em The Astrophysical Journal} {\bf 717} (2010), no.~1 1.

\bibitem{fermiarpUL}
M.~Ackermann et~al., {\it {GeV Observations of Star-forming Galaxies with the
  Fermi Large Area Telescope}},  {\em The Astrophysical Journal} {\bf 755}
  (2012), no.~2 164.

\bibitem{m82ngc253}
B.~C. {Lacki} et~al., {\it {On the GeV and TeV Detections of the Starburst
  Galaxies M82 and NGC 253}},  {\em The Astrophysical Journal} {\bf 734} (June,
  2011) 107, [\href{http://arxiv.org/abs/1003.3257}{{\tt arXiv:1003.3257}}].

\bibitem{m82nature}
{VERITAS Collaboration}, V.~A. {Acciari}, et~al., {\it {A connection between
  star formation activity and cosmic rays in the starburst galaxy M82}},  {\em
  Nature} {\bf 462} (Dec., 2009) 770--772,
  [\href{http://arxiv.org/abs/0911.0873}{{\tt arXiv:0911.0873}}].

\bibitem{ngc253hess}
A.~{Abramowski} et~al., {\it {Spectral Analysis and Interpretation of the
  {$\gamma$}-Ray Emission from the Starburst Galaxy NGC 253}},  {\em The
  Astrophysical Journal} {\bf 757} (Oct., 2012) 158,
  [\href{http://arxiv.org/abs/1205.5485}{{\tt arXiv:1205.5485}}].

\bibitem{somecomments}
D.~F. {Torres} and E.~{Domingo-Santamar{\'{\i}}a}, {\it {Some Comments on the
  High Energy Emission from Regions of Star Formation Beyond the Galaxy}},
  {\em Modern Physics Letters A} {\bf 20} (2005) 2827--2843,
  [\href{http://arxiv.org/abs/astro-ph/0509108}{{\tt astro-ph/0509108}}].

\bibitem{ArpNew}
T.~M. {Yoast-Hull}, J.~S. {Gallagher}, III, and E.~G. {Zweibel}, {\it {{Cosmic
  Rays, Gamma-Rays, \& Neutrinos in the Starburst Nuclei of Arp 220}}},  {\em
  ArXiv e-prints} (June, 2015) [\href{http://arxiv.org/abs/1506.0513}{{\tt
  arXiv:1506.0513}}].

\bibitem{arpFirst}
D.~F. {Torres}, {\it {Theoretical Modeling of the Diffuse Emission of Gamma
  Rays from Extreme Regions of Star Formation: The Case of ARP 220}},  {\em The
  Astrophysical Journal} {\bf 617} (Dec., 2004) 966--986,
  [\href{http://arxiv.org/abs/astro-ph/0407240}{{\tt astro-ph/0407240}}].

\bibitem{arphess2006}
R.~Cornils, {\em {Alignment and imaging function of the HESS reflectors and
  study of the ultraluminous infrared galaxy Arp 220 with the HESS telescope
  system}}.
\newblock PhD thesis, Universit\"at Hamburg, 2006.

\bibitem{arphess2008}
D.~Nedbal, {\em A Study of Very High Energy Gamma-Ray Emission from
  Extragalactic Objects with H.E.S.S.}
\newblock PhD thesis, Universit\"at Heidelberg, 2008.

\bibitem{magicarpUL}
J.~Albert et~al., {\it {First Bounds on the Very High Energy $\gamma$-Ray
  Emission from Arp 220}},  {\em The Astrophysical Journal} {\bf 658} (2007),
  no.~1 245.

\bibitem{whippleIC342}
A.~Merriman, {\it {{Search for very high energy Gamma radiation from the
  Starburst Galaxy IC 342}}},  Master's thesis, {{ Galway-Mayo Institute of
  Technology }}, 2010.

\bibitem{3c321maps}
D.~A. {Evans} et~al., {\it {A Radio through X-Ray Study of the
  Jet/Companion-Galaxy Interaction in 3C 321}},  {\em The Astrophysical
  Journal} {\bf 675} (2008) 1057--1066,
  [\href{http://arxiv.org/abs/0712.2669}{{\tt arXiv:0712.2669}}].

\bibitem{veritas}
J.~{Holder} et~al., {\it {Status of the VERITAS Observatory}},  in {\em
  American Institute of Physics Conference Series}, vol.~1085, pp.~657--660,
  2008.
\newblock \href{http://arxiv.org/abs/0810.0474}{{\tt arXiv:0810.0474}}.

\bibitem{naheeperformance}
N.~{Park} et~al., {\it {{Performance of the VERITAS experiment}}},  in {\em The
  34th ICRC}, 2015, in preparation.

\bibitem{veritasanalysis}
M.~Daniel, {\it {{The VERITAS standard data analysis}}},  in {\em
  {{Proceedings, 30th International Cosmic Ray Conference (ICRC 2007)}}},
  vol.~3, pp.~1325--1328, 2007.

\bibitem{trolke}
W.~A. {Rolke}, A.~M. {L{\'o}pez}, and J.~{Conrad}, {\it {Limits and confidence
  intervals in the presence of nuisance parameters}},  {\em Nuclear Instruments
  and Methods in Physics Research A} {\bf 551} (Oct., 2005) 493--503,
  [\href{http://arxiv.org/abs/physics/0403059}{{\tt physics/0403059}}].

\bibitem{cta}
M.~{Actis} et~al., {\it {Design concepts for the Cherenkov Telescope Array CTA:
  an advanced facility for ground-based high-energy gamma-ray astronomy}},
  {\em Experimental Astronomy} {\bf 32} (Dec., 2011) 193--316,
  [\href{http://arxiv.org/abs/1008.3703}{{\tt arXiv:1008.3703}}].

\end{thebibliography}\endgroup

\end{document}